\documentclass[twocolumn]{aastex631}
\usepackage{rotating}
\usepackage{amssymb}
\usepackage{amsmath}
\usepackage{color}
\usepackage{natbib}

\accepted{May 4, 2024, in ApJ Letters}
\shorttitle{SN2019ein Near a Globular Cluster}
\shortauthors{Bregman et al.}

\begin{document}

\title{A SN Ia Near a Globular Cluster in the Early-Type Galaxy NGC 5353}


\correspondingauthor{Joel N. Bregman}
\email{jbregman@umich.edu}

\author[0000-0001-6276-9526]{Joel N. Bregman}
\affiliation{Department of Astronomy \\
University of Michigan \\
Ann Arbor, MI  48109  USA}

\author[0000-0001-9852-9954]{Oleg Y. Gnedin}
\affiliation{Department of Astronomy \\
University of Michigan \\
Ann Arbor, MI  48109  USA}

\author{Patrick O. Seitzer}
\affiliation{Department of Astronomy \\
University of Michigan \\
Ann Arbor, MI  48109  USA}

\author{Zhijie Qu}
\affiliation{Department of Astronomy and Astrophysics\\
University of Chicago \\
Chicago, IL 60637  USA}

\begin{abstract}

No progenitor of a Type Ia supernova is known, but in old population early-type galaxies, one may find SN Ia associated with globular clusters, yielding a population age and metallicity.  It also provides insight into the formation path and the SN enhancement rate in globular clusters. We sought to find such associations and identified SN 2019ein to be within the ground-based optical positional uncertainty of a globular cluster candidate within the early-type galaxy NGC 5353 at D $\approx 30$ Mpc.  We reduced the positional uncertainties by obtaining \textit{Hubble Space Telescope} images with the \textit{Advanced Camera for Surveys}, using filters F475W and F814W and obtained in June 2020.  We find that the globular cluster candidate has a magnitude, color, and angular extent that are consistent with it being a typical globular cluster.  The separation between the globular cluster and SN 2019ein is 0.43$^{\prime \prime}$, or 59 pc in projection.  The chance occurrence with a random globular cluster is $\approx 3$\%, favoring but not proving an association.  If the SN progenitor originated in the globular cluster, one scenario is that SN 2019ein was previously a double degenerate white dwarf binary that was dynamically ejected from the globular cluster and exploded within 10 Myr; models do not predict this to be common.  Another, but less likely scenario is where the progenitor remained bound to the globular cluster, allowing the double degenerate binary to inspiral on a much longer timescale before producing a SN. 

\end{abstract}


\section{Introduction} \label{sec:intro}

Type Ia SNe have become special objects in astronomy, partly because of their central 
role as standard candles that are visible at cosmological distances.  It was with 
these objects that the acceleration of the universe was first detected and they will 
play an important role in accurately measuring the acceleration parameters over cosmological time.  
This greater accuracy requires not only a large number of supernovae, but a good 
understanding of the supernova event.  
However, Ia progenitors have not been identified nor is the exact model leading to the explosion understood (e.g., \citealt{Maoz2014, Branch2017, Soker2019, LiuSNIaRev2023}).

There is agreement that SN Ia occurs from binary systems where one or both stars are white dwarfs (the single-degenerate, SD, or the double-degenerate, DD models).  There are several subclasses of SN Ia, as seen in a peak magnitude vs decline rate plot \citep{Taubenberger2017}, but about 70\% of SNIa are homogeneous in that there is an excellent relationship between the width of the light curve and the peak optical luminosity \citep{Phillips1999}, making them suitable as standard candles.  

Despite intense study of ``normal'' SN Ia, we are not certain of the likely progenitor system or the particular explosion scenario.  There is no detection of the progenitor systems or of a post-SN surviving binary, even for some of the nearer galaxies.  For the explosion mechanism, the ``classical'' models posited that the mass of the white dwarf (WD) is pushed over the Chandrasekhar limit, either from Roche Lobe overflow from a star (SD), or through accretion from a companion WD (DD).  Both situations probably occur, but it is unclear if either can produce the most normal SN Ia.  There are issues with each channel, such as for the SD case, where the accretion should be visible in X-rays, but too few such soft X-ray emitters are detected \citep{Gilfanov2010}.  

Another model that has received broad attention is where the WD has a sub-Chandrasekhar mass of $\approx 1.0$ M$_{\odot}$ (e.g., \citealt{Shen2018}).  The WD accretes He from a companion, such as a He star or He WD, and when the He mass shell grows to $\sim 0.02$ M$_{\odot}$, a He-shell detonation event occurs.  This first detonation compresses the carbon core of the WD, causing a second detonation that is responsible for the SN Ia. There are variations on this double-detonation model such as involving violet mergers, for example (see \citealt{LiuSNIaRev2023} for a comprehensive review).

A prerequisite for the formation of a close binary that may become a SN Ia is that the two stars have a small enough separation that they interact through a common envelope or Roche Lobe mass transfer.  This separation should be less than $\sim 1$ AU, which is the approximate radius of a 1-3 M$_{\odot}$ during the AGB stage. 
However, most binary stars of types F to G in the Solar neighborhood have a typical separation of $\sim 30$ AU \citep{Duquennoy1991}, so very few become close enough to interact through mass exchange. 
Any mechanism that can reduce the separation of binary stars will increase the SN Ia rate.
Binary-binary encounters can shrink the orbital separation, but the only systems with dynamical relaxation times less than a Hubble time are dense stellar clusters \citep{Shara2002}, such as globular clusters (GC).

One might think that the observation that SN Ia is seen in early-type galaxies proves that
they occur in old populations.  However, there is a small amount of star formation in early-type
galaxies and the small number of A and F stars may be responsible for some SN Ia.  
There are a few examples where this is the case, such as SN 1994d in NGC 4526, where it
occurs on the edge of a rather modest disk of gas and dust \citep{Koku2017,Nyland2017}.  
Such disks are not terribly uncommon in early-type galaxies and \cite{Wash2013} 
find that about 1/3 of SN Ia's are clearly associated with an extinction region, such as a disk. 
Even in cases where a SN Ia occurred in a nearby galaxy, such as the
event in M 101 (SN 2011fe; D = 6.9 Mpc), the progenitor is not visible in the many \textit{HST} observations during the years leading to the event \citep{Weidong2011,Bloom2012}.

One can avoid confusing these young SN Ia with older ones when the SN lies in a globular cluster.  
Nearly all globular clusters are older than 3 Gyr and it is easy 
to identify the small fraction of globular clusters that are younger.
Most globular cluster stars have nearly the same age and approximately the same [Fe/H].
Knowing progenitor properties, such as the age and [Fe/H], places important constraints
on models \citep{Koba1998,Pfahl2009}, thereby restricting the possible 
parameter space (e.g., \citealt{Timmes2003, Scalzo2019}.
Showing that a SN Ia occurred in a globular cluster would be definitive support for the double degenerate path.

The rate of low-mass X-ray binaries (usually a star orbiting a neutron star) is enhanced in globular clusters by a factor of about 200 (discussion in \citealt{Wash2013}).  If this enhancement were to apply to double white dwarf binaries, it might imply a similar enhancement for the SN Ia rate.
This possibility motivated two studies that investigated the association between GCs and SN Ia by examining archival \textit{HST} imaging data of nearby early-type galaxies in which SN Ia events occurred in the past few decades \citep{Voss2012,Wash2013}.
This is only possible for galaxies within $\sim 30$ Mpc, so that a modest depth \textit{HST}
image would detect globular clusters.
The images were taken for other purposes before the SN event or a few years after
the event, when the SN faded below the magnitudes of globular clusters.
Many of the events lacked any \textit{HST} images, and for the existing images, there was a strong
bias toward coverage of the inner part of the galaxy.  There were challenges 
with working with images of the inner parts of galaxies:  the surface brightness can
be high, making it difficult to detect GCs; and there are many GCs in the inner part of galaxies, leading to frequent chance coincidences.
Of the SN Ia events examined, there are some ambiguous cases (low S/N; high probability of chance coincidence) but neither study found a single definitive case of an association between a GC and a SN Ia.
Both studies concluded that the enhancement rate for GCs producing SN Ia was $< 50$, 
compared to the field star production rate. 
There is also the possibility that the SN Ia progenitor was created in the globular cluster but ejected from it before the SN Ia event occurred (considered in $\S$4). 

Here we seek to identify any recent SN Ia with globular clusters in nearby galaxies (D $\lesssim 30$ Mpc), preferably with a pre-SN \textit{HST} image, so that one can determine if a globular cluster candidate is close to the SN.
There are generally fewer than one such event per year.
There are additional considerations, such as avoiding the central region where the density of globular clusters is high, which can lead to false positives.  

An ideal case is found with SN 2019ein, in the early-type system NGC 5353 (D = 27.8 Mpc), and far enough from the center (5.5 kpc) that the projected density of globular clusters is not high.
An \textit{HST} image of NGC 5353, including the location of SN 2019ein. was obtained four years before the SN event (WFC3 IR, F110W).
It showed a globular cluster candidate that was within 1$\arcsec$ of SN 2019ein, the uncertainty mainly due to the positional error in the ground-based images. 
However, 1$\arcsec$ corresponds to 135 pc, which is much larger than the typical size of a globular cluster ($\lesssim 20$ pc, or $0.15\arcsec$).  
One can improve greatly on the position of the SN relative to the globular cluster by obtaining an image containing SN 2019ein.  A set of optical images (ACS F475W and F814W) was obtained in June 2020, when SN 2019ein was easily visible but not vastly brighter than the globular cluster counterpart.
Here, we report on the observations and discuss the conditions under which SN 2019ein could be related to the potential globular cluster.

\begin{figure}
\plotone{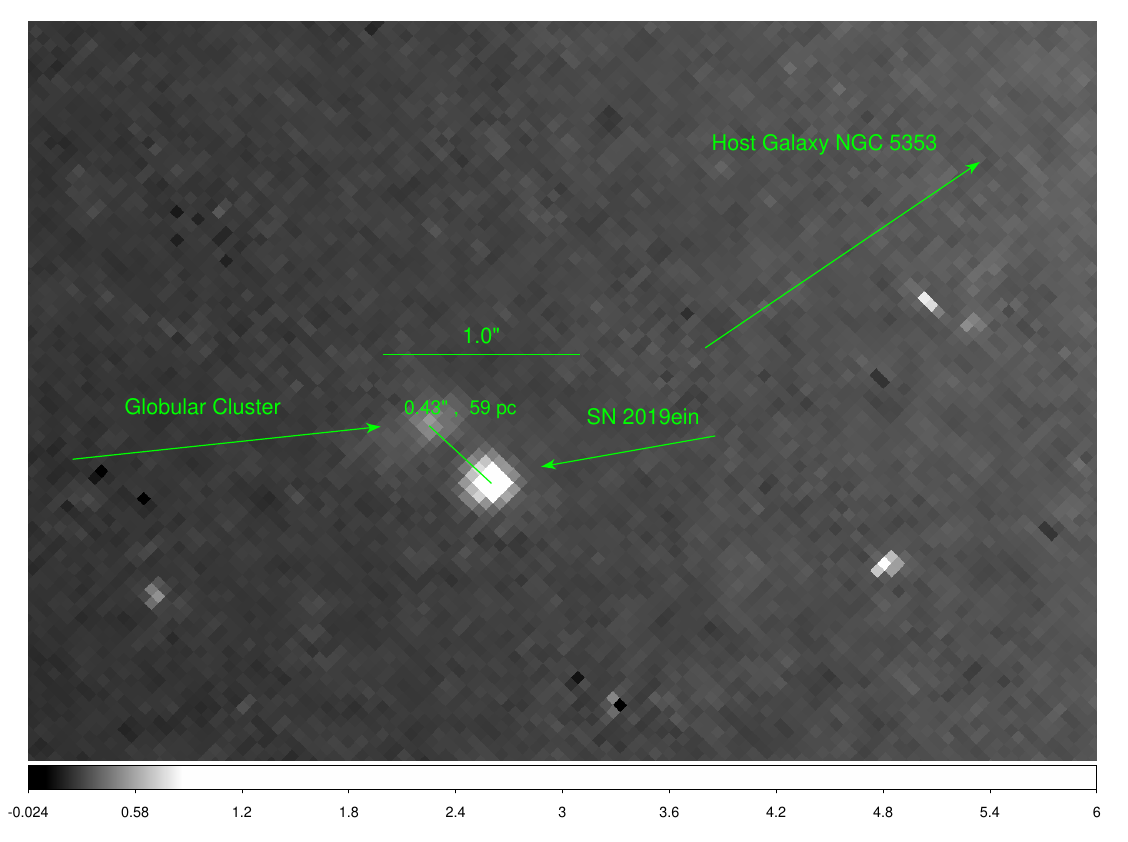}
\caption{The \textit{HST}/ACS/F814W image of the region around SN 2019ein, taken in Jun 2020, with the globular cluster candidate and the supernova marked; the center of NGC 5353 is to the NW.  The globular cluster candidate is larger than the point spread function, which is typical for an average globular cluster at this distance. 
\label{fig:fig1}
}
\end{figure}

\section{Observations}

SN 2019ein was discovered on 1 May 2019 (by the ATLAS program; \citealt{SN2019einDiscovery}) with a brightness of 13.9 mag and lies 
about 5.5 kpc (40$^{\prime\prime}$) from the center of NGC 5353.  It has since been observed
many times for both photometry and spectroscopy \citep{Kawa2020}, so the type is well-established as Type Ia.
The light curve for a typical SN Ia \citep{Graham2017,Scalzo2019} predicted that SN 2019ein would be easily visible with \textit{HST} imaging through 2020.  
It was observed through program 16047 with both the F475W and F814W filters on the ACS/WFC.
The first observing attempt, in March 2020, failed due to an inability to lock on to a suitable guide star, but this issue was resolved in a successful set of observations on June 8, 2020.
The target was positioned toward the end of the camera so that a subarray could be used, thereby improving the efficiency and  minimizing the charge transfer efficiency (CTE) issue (using the WFC1-CTE aperture). 
A two-point linear dithering pattern was used, with a small step that improved the PSF. 
Excellent images were obtained in the two filters, each with an exposure time of 1064 sec. 

In addition to these two observations, there was an archival IR observation of the field containing SN 2019ein, obtained prior to the SN.  This observation was obtained on December 19, 2015 with the F110W on the Wide Field Camera 3 (WFC3/IR), with an exposure time of 1997 sec. 
We used this image to originally identify the possible association of SN 2019ein with a globular cluster candidate. 
This observation suggested an object that is not a point source and has a F110W magnitude of 23.73.  

For the June 2020 images, standard data processing was used. In particular, we analyzed the calibrated, geometrically-corrected, dither-combined image (created by AstroDrizzle), along with the CTE correction (the drc.fits image).  
From the images, we obtained relative positions, magnitudes, and shapes for comparison with the PSF.
Without correcting the image coordinates to the GAIA data, the center of the globular cluster candidate is RA, DEC (J2000) = 13:53:29.141, +40:16:31.71, and SN 2019ein lies 0.43'' to the SW (at RA, DEC of 13:53:29.202,  +40:16:29.55).    

For the globular cluster candidate, m(g) = 25.64 $\pm 0.10$, m(F814W) = 24.01 $\pm 0.1$, and m(F110W) = 23.72 $\pm 0.04$ (note that the F110W filter is known as Wide YJ, with a center at 1.153 $\mu$m). 
Galactic extinction is modest at this high Galactic latitude (\textit{b} = 71.63$^{\circ}$), with A(g) = 0.042, A(F814W) = 0.02, and A(F110W) = 0.011.
The colors of the globular cluster, uncorrected for extinction are m(g) - m(F814W) = 1.63 $\pm$ 0.12 and m(F814W) - m(F110W) = 0.29.
For a distance of 27.8 Mpc, m-M =  32.22, so the absolute magnitudes, corrected for extinction are M(g) = -6.62$\pm 0.10$, M(F814W) = -8.23$\pm 0.10$, and M(F110W) = -8.51$\pm 0.04$, with colors m(g) - m(F814W) = 1.61 $\pm$ 0.13 and m(F814W) - m(F110W) = 0.28 $\pm$ 0.11.  
These are the statistical magnitude uncertainties, but there is also about a 10\% uncertainty in the distance, which should be considered. For example, some distances are greater by 0.45 magnitudes (the Extragalactic Distance Database \citealt{Tully2009}). Then, the globular cluster candidate would be M(F814W) = -8.68, one magnitude brighter than the typical peak, but well within the distribution \citep{Harris2014LF}.
For the larger distance, the angular to linear sizes increase by 23\%. 

The observed magnitudes of SN 2019ein on June 8, 2020 was m(F475W) = m(g) = 22.61$\pm 0.06$ and m(F814W) = 22.51$\pm 0.06$ (the F814W filter is referred to as Broad I, and is approximately equivalent to Johnson I for objects of this color).  The color is m(g) - m(F814W) = 0.10$\pm 0.08$, or corrected for extinction, 0.08$\pm 0.08$. 
The magnitudes of SN 2019ein decline are approximately those expected for a SN Ia, 
relative to earlier observations \citep{Gaobo2022}. 

\section{The Globular Cluster Candidate and the Association with SN 2019ein}

The globular cluster candidate is too faint to obtain a spectrum, so we consider its properties from its absolute magnitude, colors, and whether it is extended beyond the ACS point spread function.  The absolute magnitude, M(F814W) = -8.23 (-8.68 for the longer distance; logL$_*$/L$_\odot$ = 4.95, 5.13 respectively), which can be compared to the turnover magnitude and color of a globular cluster population in an early type galaxy.  The turnover magnitude rises with increasing galaxy luminosity \citep{Harris2014LF}, and for the luminosity of NGC 5353, the turnover luminosity is M(F814W) $\approx -8.4$, which is close to that for M87 of M(F814W) = -8.56 \citep{Peng2009}; the width of these distributions is $\approx 1.3$ mag.  Thus the absolute magnitude of the globular cluster candidate is near the turnover in the luminosity function and and within the width of the distribution for either choice of distance to NGC 5353.  Also, the color M(g) - M(F814W) = 1.61 is typical for an average globular cluster, where the observed range is 1-2.3 (e.g., \citealt{Harris2016N6166}).  This color is at the location where the contributions from an old and young population are similar \citep{Harris2014LF}.

A typical globular cluster half-light radius is $\sim 3$ pc (0.02$^{\prime \prime}$ at the distance of NGC 5353), so the inner part of a cluster would be unresolved to the ACS.  However, the tidal radius is often an order of magnitude larger (e.g., \citealt{Moreno2014}), so for globular clusters in a galaxy at the distance of NGC 5353, they are more extended than point sources.  
We can see that the globular cluster candidate has a FWHM about twice that of a point source, such as the nearby supernova.  This is consistent with the departure from a point source that is expected for a globular cluster at this distance, but the S/N is not adequate to determine a value for the half-light radius. There is a hint of an extension of light to the SE, but it is of a low statistical significance. 
The magnitude, color, and angular size are all consistent with an average globular cluster in NGC 5353. 

We considered the chance probability that the globular cluster candidate is a background galaxy and find that unlikely.  For its m(814W) value, the sky density of objects is about 20-30 arcmin$^{-2}$, based on the Hubble Ultra Deep Field observations \citep{Beckwith2006}.  The chance of finding any object within 0.5'' of SN2019ein is about 1/600.  
This chance probability rate decreases when we include the constraints that it have the observed color, that it has a small projected angular size that is consistent with being spherical.  
These considerations would reduce the chance that it is a background galaxy to well below 0.1\%.

The number of globular clusters in a galaxy is correlated with the luminosity of the system (e.g., \citealt{Eadie2022}), although there is scatter at a given stellar mass of the galaxy.  
From the compilations of globular cluster number vs stellar mass, or absolute IR magnitude \citep{Eadie2022}, we estimate the number of of globular clusters to be $\approx 500$ in NGC 5353, using its near IR magnitude (H = -24.50).
The globular cluster radial distribution approximately follows the galaxy surface brightness.  This distribution was fit to other early-type galaxies using a beta model, where the projected number density of globular clusters follows $N(r) = N_0 [1+(r/r_c)^2]^{-3\beta+0.5}$, and with an average value of $\beta = 0.6$ \citep{Forbes2012}.
The normalization $N_0$ is chosen so that the total integrated number of globular clusters is 500, or $N_0 = 5.76$ globular clusters per arcsec$^2$.  
We use $r_c = 500$ pc, but the results are not very sensitive to this value as long as it lies within the range of observed values. 
At the location of SN 2019ein, we calculate that N = 0.043 gc arcsec$^{-2}$.  
The circular area defined by the separation between the globular cluster candidate and SN 2019ein is 0.58 arcsec$^2$, so the number of globular clusters within that circle is 0.025.  That is, there is a probability of associating the supernova with the globular cluster candidate of about 3\%. 

We can check the projected number density by counting the number of globular cluster candidates at the same separation from the center of NGC 5353, but on the mirror side of the major axis. 
We did this within a 10$^{\prime \prime}$ radius and identified all objects that were more extended than the PSF but were not obviously a much larger background galaxy.  There were 15 objects in that circle that fit the above criteria and also had a m(F475W) - m(F814W) color in the range found for globular clusters.  This leads to a projected density of 0.048 gc arcsec$^{-2}$, which is close to the above result. 
We conclude that it is likely, but not certain for SN 2019ein to be associated with the globular cluster candidate.

\section{Interpretation and Discussion}

\subsection{Is SN2019ien Bound to the Globular Cluster?}

If SN 2019ein is not physically associated with the globular cluster candidate, there is no need for an interpretation of the observational result.  In the following, we consider the implications that result if SN 2019ein is or was associated with the globular cluster candidate, which we will assume is a globular cluster in NGC 5353.

An important issue is how far from the globular cluster can a star still be gravitationally bound. 
This requires that we know the tidal radius of the globular cluster.
To accurately make this calculation, one needs to know the orbital parameters of the globular cluster, its mass and half-light radius, as well as the axisymmetric and non-axisymmetric potential components of NGC 5353.  
We lack nearly all of this information, and the only galaxy in which tidal radii have been calculated for individual globular clusters is in the Milky Way \citep{Moreno2014}.  
From this work, and for globular clusters not near the center of the Milky Way ($r > 3$ kpc) the typical tidal radius is 20-100 pc, with a median value of about 40 pc.  This can be compared to the projected separation of SN 2019ein to the globular cluster of 59 pc.  This represents the minimum separation as the true physical separation is larger by a factor of 1/$sin(i)$, where $i$ is the angle between our line of sight and a vector connecting the globular cluster and SN 2019ein.
The average correction for a large number of randomly oriented vectors would raise the separation to $\sim 100$ pc.  
These estimates suggest that SN 2019ein probably does not lie within the tidal radius of the globular cluster candidate, but without orbital information, this is not a strong conclusion. 

\subsection{Theoretical and Observational Constraints on the WD Binary System}

The condition that a SN Ia progenitor be ejected (or nearly ejected) from a globular cluster places very particular conditions on the nature of the WD binary.  For this discussion, we assume that the globular cluster has evolved so that the ``burning'' of black hole binaries is mostly complete and that the burning of WD binaries is occurring.  The interactions between hard binaries can lead to an increase in the orbital velocity through the globular cluster, at the expense of the orbital velocity of the secondary around the primary.  When this velocity increase is comparable to or greater than the velocity dispersion of the cluster, the hard binary is cast out of the core if not the entire cluster.  This is the mechanism by which black holes are ejected from globular clusters (e.g., \citealt{Aar2012, Kremer2020ApJS}).  

The ejection of WD binaries occurs less frequently because the internal binary velocities are lower \citep{Kremer2021}.  For example, a binary between a WD and the He core of a star has a typical orbital velocity of 30 km s$^{-1}$, but the transfer to the orbital motion of the binary through the cluster is at the $\sim10$\% level.  This would not lead to a significant increase in the semi-major axis of the orbit of the binary through the GC. 
The situation is more favorable for a WD-WD binary, where the internal orbital velocity is $\lesssim 120$ km s$^{-1}$ for two 1 M${_\odot}$ WD.  The velocity added to the orbital velocity through the globular cluster might occasionally send the binary far from the core.  This is the only scenario that would be consistent with the observed situation for SN2019ein. 
The possible SNe mechanisms are for the merger of the two WD, or for accretion from the secondary onto the primary, followed by a double detonation event. 

The analysis of the observations of SN2019ein further limits the possibilities for the WD binary and the nature of the explosion. 
From light curves and spectra taken in the first year \citep{Kawabata2020}, SN2019ein seems normal and consistent with the 1D delayed detonation model, although they also consider a sub-Chandrasekhar mass WD model to be viable.  
A spectropolarimetry effort found very low optical polarization \citep{Patra2022}, 0-0.3\%, in the first month after the explosion, which rules out significant departures from spherical symmetry.  They argue that this is evidence against the merger-induced and double-detonation models.
From a longer timeline of photometric and spectroscopic observations (3-330 days), \cite{Gaobo2022} show that SN2019ein is about 40\% fainter than a normal SNe Ia, with a correspondingly lower nickel mass.  They argue that a double-detonation mechanism was responsible, with a WD mass of $M_{WD} \approx 1 M_{\odot}$, and a low mass He shell (0.01 $M_{\odot}$).
For consistency with the above dynamical argument, the secondary could be a He WD and the primary would be a sub-Chandrasekhar mass WD.  Then, the SNe occurs through the double detonation model (this is not the only possibility; see \citealt{LiuSNIaRev2023}).  

\subsection{The Last 10 Myr Before SN2019ein}

The picture outlined above is that there was a binary-binary interaction that led the SN2019ein progenitor to escape the core before becoming a SNe.  This implies a delay time between the dual binary interaction and the SNe.  For a characteristic kick velocity of 30 km s$^{-1}$, the time needed to reach the position of the SNe is $\sim 3-10 \, (v/30$ km s$^{-1}$)$^{-1}$ Myr. 
So, after the binary interaction the secondary WD did not quite fill its Roche Lobe, but had to inspiral until it was close enough to initiate mass transfer or merge.  
The inspiral time distribution for ejected WD binaries was calculated by 
\cite{Kremer2020ApJS}, who find that about 10\% of such objects have merger times less than 10 Myr. 
The mass transfer time to build up 0.01 M$_{\odot}$ of He on the surface of the primary WD is somewhat shorter, typically 1-2 Myr. 

\section{Summary}

A SN Ia, SN 2019ein, was found to lie close to a globular cluster candidate in the early-type galaxy NGC 5353, based on a pre-SN near-infrared image with the WFC on \textit{HST}. 
In our subsequent program with the ACS taken in June 2020, both the F475W and F814W images show SN 2019, which is separated from the globular cluster candidate by 0.43$^{\prime \prime}$ (59 pc). 
The color, magnitude, and angular extent of the globular cluster candidate is consistent with being an average globular cluster in an early-type galaxy.
The chance coincidence of SN 2019ein with a globular cluster 5.5 kpc from the center of NGC 5353 is approximately 3\%, so a true association between the two objects appears likely (but not definitive). 

Assuming that SN 2019ein is or was physically associated with the globular cluster, it would probably be unbound from the globular cluster, given the separation. 
In the case that it was previously bound to the globular cluster core, there would have been an ejection of a double white dwarf binary at $\sim 30$ km s$^{-1}$.
The WD binary would then have had to inspiral within about 10 Myr of the ejection event to either merge or produce Roche lobe overflow of He onto the sub-Chandrasekhar mass WD.
This would be a rare but possible event, based on the fraction of WD binary ejections and the relatively short inspiral time required (\citealt{Kremer2021}), but see \citealt{Ivanova2006} for a higher rate).  
Less likely but not completely ruled out, due to lack of dynamical information, is a scenario where the double degenerate binary remained bound to the globular cluster. 
This would permit inspiral times to be much greater than 10 Mpc. 
We note that an association of SN 2019ein with a globular cluster is consistent with the limits on the SN Ia enhancement rates of \cite{Voss2012} and \cite{Wash2013}.   

\section*{Acknowledgements}

We thank the many individuals who offered guidance and insight, including Kyle Kremer, Noam Soker, Fred Rasio, Cameron Pratt, and Anne Blackwell.
We are particularly grateful to Shelly Meyett and Samantha Hoffmann at STScI for their aid and patience in planning the observations. 
We are grateful for support from NASA and STScI through the program and award HST-GO-16047.
This research has made use of the NASA/IPAC Extragalactic Database (NED),
which is operated by the Jet Propulsion Laboratory, California Institute of Technology, under contract with NASA.

All of the data presented in this article were obtained from the Mikulski Archive for Space Telescopes (MAST) at the Space Telescope Science Institute. The specific observations analyzed can be accessed via \dataset[DOI: 10.17909/1375-6a79]{https://doi.org/10.17909/1375-6a79}.
\newpage

\bibliography{SN2019ein}{}

\begin{thebibliography}{}
\expandafter\ifx\csname natexlab\endcsname\relax\def\natexlab#1{#1}\fi
\providecommand{\url}[1]{\href{#1}{#1}}
\providecommand{\dodoi}[1]{doi:~\href{http://doi.org/#1}{\nolinkurl{#1}}}
\providecommand{\doeprint}[1]{\href{http://ascl.net/#1}{\nolinkurl{http://ascl.net/#1}}}
\providecommand{\doarXiv}[1]{\href{https://arxiv.org/abs/#1}{\nolinkurl{https://arxiv.org/abs/#1}}}

\bibitem[{{Aarseth}(2012)}]{Aar2012}
{Aarseth}, S.~J. 2012, \mnras, 422, 841, \dodoi{10.1111/j.1365-2966.2012.20666.x}

\bibitem[{{Beckwith} {et~al.}(2006){Beckwith}, {Stiavelli}, {Koekemoer}, {Caldwell}, {Ferguson}, {Hook}, {Lucas}, {Bergeron}, {Corbin}, {Jogee}, {Panagia}, {Robberto}, {Royle}, {Somerville}, \& {Sosey}}]{Beckwith2006}
{Beckwith}, S. V.~W., {Stiavelli}, M., {Koekemoer}, A.~M., {et~al.} 2006, \aj, 132, 1729, \dodoi{10.1086/507302}

\bibitem[{{Bloom} {et~al.}(2012){Bloom}, {Kasen}, {Shen}, {Nugent}, {Butler}, {Graham}, {Howell}, {Kolb}, {Holmes}, {Haswell}, {Burwitz}, {Rodriguez}, \& {Sullivan}}]{Bloom2012}
{Bloom}, J.~S., {Kasen}, D., {Shen}, K.~J., {et~al.} 2012, \apjl, 744, L17, \dodoi{10.1088/2041-8205/744/2/L17}

\bibitem[{{Branch} \& {Wheeler}(2017)}]{Branch2017}
{Branch}, D., \& {Wheeler}, J.~C. 2017, {Supernova Explosions}, \dodoi{10.1007/978-3-662-55054-0}

\bibitem[{{Duquennoy} \& {Mayor}(1991)}]{Duquennoy1991}
{Duquennoy}, A., \& {Mayor}, M. 1991, \aap, 248, 485

\bibitem[{{Eadie} {et~al.}(2022){Eadie}, {Harris}, \& {Springford}}]{Eadie2022}
{Eadie}, G.~M., {Harris}, W.~E., \& {Springford}, A. 2022, \apj, 926, 162, \dodoi{10.3847/1538-4357/ac33b0}

\bibitem[{{Forbes} {et~al.}(2012){Forbes}, {Ponman}, \& {O'Sullivan}}]{Forbes2012}
{Forbes}, D.~A., {Ponman}, T., \& {O'Sullivan}, E. 2012, \mnras, 425, 66, \dodoi{10.1111/j.1365-2966.2012.21368.x}

\bibitem[{{Gilfanov} \& {Bogd{\'a}n}(2010)}]{Gilfanov2010}
{Gilfanov}, M., \& {Bogd{\'a}n}, {\'A}. 2010, \nat, 463, 924, \dodoi{10.1038/nature08685}

\bibitem[{{Graham} {et~al.}(2017){Graham}, {Kumar}, {Hosseinzadeh}, {Hiramatsu}, {Arcavi}, {Howell}, {Valenti}, {Sand}, {Parrent}, {McCully}, \& {Filippenko}}]{Graham2017}
{Graham}, M.~L., {Kumar}, S., {Hosseinzadeh}, G., {et~al.} 2017, \mnras, 472, 3437, \dodoi{10.1093/mnras/stx2224}

\bibitem[{{Harris} {et~al.}(2016){Harris}, {Blakeslee}, {Whitmore}, {Gnedin}, {Geisler}, \& {Rothberg}}]{Harris2016N6166}
{Harris}, W.~E., {Blakeslee}, J.~P., {Whitmore}, B.~C., {et~al.} 2016, \apj, 817, 58, \dodoi{10.3847/0004-637X/817/1/58}

\bibitem[{{Harris} {et~al.}(2014){Harris}, {Morningstar}, {Gnedin}, {O'Halloran}, {Blakeslee}, {Whitmore}, {C{\^o}t{\'e}}, {Geisler}, {Peng}, {Bailin}, {Rothberg}, {Cockcroft}, \& {Barber DeGraaff}}]{Harris2014LF}
{Harris}, W.~E., {Morningstar}, W., {Gnedin}, O.~Y., {et~al.} 2014, \apj, 797, 128, \dodoi{10.1088/0004-637X/797/2/128}

\bibitem[{{Ivanova} {et~al.}(2006){Ivanova}, {Heinke}, {Rasio}, {Taam}, {Belczynski}, \& {Fregeau}}]{Ivanova2006}
{Ivanova}, N., {Heinke}, C.~O., {Rasio}, F.~A., {et~al.} 2006, \mnras, 372, 1043, \dodoi{10.1111/j.1365-2966.2006.10876.x}

\bibitem[{{Kawabata} {et~al.}(2020{\natexlab{a}}){Kawabata}, {Maeda}, {Yamanaka}, {Nakaoka}, {Kawabata}, {Adachi}, {Akitaya}, {Burgaz}, {Hanayama}, {Horiuchi}, {Hosokawa}, {Iida}, {Imazato}, {Isogai}, {Jiang}, {Katoh}, {Kimura}, {Kino}, {Kuroda}, {Maehara}, {Matsubayashi}, {Morihana}, {Murata}, {Nagao}, {Niwano}, {Nogami}, {Oeda}, {Ono}, {Onozato}, {Otsuka}, {Saito}, {Sasada}, {Shiraishi}, {Sugiyama}, {Taguchi}, {Takahashi}, {Takagi}, {Takagi}, {Takayama}, {Tozuka}, \& {Sekiguchi}}]{Kawa2020}
{Kawabata}, M., {Maeda}, K., {Yamanaka}, M., {et~al.} 2020{\natexlab{a}}, \apj, 893, 143, \dodoi{10.3847/1538-4357/ab8236}

\bibitem[{{Kawabata} {et~al.}(2020{\natexlab{b}}){Kawabata}, {Maeda}, {Yamanaka}, {Nakaoka}, {Kawabata}, {Adachi}, {Akitaya}, {Burgaz}, {Hanayama}, {Horiuchi}, {Hosokawa}, {Iida}, {Imazato}, {Isogai}, {Jiang}, {Katoh}, {Kimura}, {Kino}, {Kuroda}, {Maehara}, {Matsubayashi}, {Morihana}, {Murata}, {Nagao}, {Niwano}, {Nogami}, {Oeda}, {Ono}, {Onozato}, {Otsuka}, {Saito}, {Sasada}, {Shiraishi}, {Sugiyama}, {Taguchi}, {Takahashi}, {Takagi}, {Takagi}, {Takayama}, {Tozuka}, \& {Sekiguchi}}]{Kawabata2020}
---. 2020{\natexlab{b}}, \apj, 893, 143, \dodoi{10.3847/1538-4357/ab8236}

\bibitem[{{Kobayashi} {et~al.}(1998){Kobayashi}, {Tsujimoto}, {Nomoto}, {Hachisu}, \& {Kato}}]{Koba1998}
{Kobayashi}, C., {Tsujimoto}, T., {Nomoto}, K., {Hachisu}, I., \& {Kato}, M. 1998, \apjl, 503, L155, \dodoi{10.1086/311556}

\bibitem[{{Kokusho} {et~al.}(2017){Kokusho}, {Kaneda}, {Bureau}, {Suzuki}, {Murata}, {Kondo}, \& {Yamagishi}}]{Koku2017}
{Kokusho}, T., {Kaneda}, H., {Bureau}, M., {et~al.} 2017, \aap, 605, A74, \dodoi{10.1051/0004-6361/201630158}

\bibitem[{{Kremer} {et~al.}(2021){Kremer}, {Rui}, {Weatherford}, {Chatterjee}, {Fragione}, {Rasio}, {Rodriguez}, \& {Ye}}]{Kremer2021}
{Kremer}, K., {Rui}, N.~Z., {Weatherford}, N.~C., {et~al.} 2021, \apj, 917, 28, \dodoi{10.3847/1538-4357/ac06d4}

\bibitem[{{Kremer} {et~al.}(2020){Kremer}, {Ye}, {Rui}, {Weatherford}, {Chatterjee}, {Fragione}, {Rodriguez}, {Spera}, \& {Rasio}}]{Kremer2020ApJS}
{Kremer}, K., {Ye}, C.~S., {Rui}, N.~Z., {et~al.} 2020, \apjs, 247, 48, \dodoi{10.3847/1538-4365/ab7919}

\bibitem[{{Li} {et~al.}(2011){Li}, {Bloom}, {Podsiadlowski}, {Miller}, {Cenko}, {Jha}, {Sullivan}, {Howell}, {Nugent}, {Butler}, {Ofek}, {Kasliwal}, {Richards}, {Stockton}, {Shih}, {Bildsten}, {Shara}, {Bibby}, {Filippenko}, {Ganeshalingam}, {Silverman}, {Kulkarni}, {Law}, {Poznanski}, {Quimby}, {McCully}, {Patel}, {Maguire}, \& {Shen}}]{Weidong2011}
{Li}, W., {Bloom}, J.~S., {Podsiadlowski}, P., {et~al.} 2011, \nat, 480, 348, \dodoi{10.1038/nature10646}

\bibitem[{{Liu} {et~al.}(2023){Liu}, {R{\"o}pke}, \& {Han}}]{LiuSNIaRev2023}
{Liu}, Z.-W., {R{\"o}pke}, F.~K., \& {Han}, Z. 2023, Research in Astronomy and Astrophysics, 23, 082001, \dodoi{10.1088/1674-4527/acd89e}

\bibitem[{{Maoz} {et~al.}(2014){Maoz}, {Mannucci}, \& {Nelemans}}]{Maoz2014}
{Maoz}, D., {Mannucci}, F., \& {Nelemans}, G. 2014, \araa, 52, 107, \dodoi{10.1146/annurev-astro-082812-141031}

\bibitem[{{Moreno} {et~al.}(2014){Moreno}, {Pichardo}, \& {Vel{\'a}zquez}}]{Moreno2014}
{Moreno}, E., {Pichardo}, B., \& {Vel{\'a}zquez}, H. 2014, \apj, 793, 110, \dodoi{10.1088/0004-637X/793/2/110}

\bibitem[{{Nyland} {et~al.}(2017){Nyland}, {Young}, {Wrobel}, {Davis}, {Bureau}, {Alatalo}, {Morganti}, {Duc}, {de Zeeuw}, {McDermid}, {Crocker}, \& {Oosterloo}}]{Nyland2017}
{Nyland}, K., {Young}, L.~M., {Wrobel}, J.~M., {et~al.} 2017, \mnras, 464, 1029, \dodoi{10.1093/mnras/stw2385}

\bibitem[{{Patra} {et~al.}(2022){Patra}, {Yang}, {Brink}, {H{\"o}flich}, {Wang}, {Filippenko}, {Kasen}, {Baade}, {Foley}, {Maund}, {Zheng}, {Hung}, {Cikota}, {Wheeler}, \& {Bulla}}]{Patra2022}
{Patra}, K.~C., {Yang}, Y., {Brink}, T.~G., {et~al.} 2022, \mnras, 509, 4058, \dodoi{10.1093/mnras/stab3136}

\bibitem[{{Peng} {et~al.}(2009){Peng}, {Jord{\'a}n}, {Blakeslee}, {Mieske}, {C{\^o}t{\'e}}, {Ferrarese}, {Harris}, {Madrid}, \& {Meurer}}]{Peng2009}
{Peng}, E.~W., {Jord{\'a}n}, A., {Blakeslee}, J.~P., {et~al.} 2009, \apj, 703, 42, \dodoi{10.1088/0004-637X/703/1/42}

\bibitem[{{Pfahl} {et~al.}(2009){Pfahl}, {Scannapieco}, \& {Bildsten}}]{Pfahl2009}
{Pfahl}, E., {Scannapieco}, E., \& {Bildsten}, L. 2009, \apjl, 695, L111, \dodoi{10.1088/0004-637X/695/1/L111}

\bibitem[{{Phillips} {et~al.}(1999){Phillips}, {Lira}, {Suntzeff}, {Schommer}, {Hamuy}, \& {Maza}}]{Phillips1999}
{Phillips}, M.~M., {Lira}, P., {Suntzeff}, N.~B., {et~al.} 1999, \aj, 118, 1766, \dodoi{10.1086/301032}

\bibitem[{{Scalzo} {et~al.}(2019){Scalzo}, {Parent}, {Burns}, {Childress}, {Tucker}, {Brown}, {Contreras}, {Hsiao}, {Krisciunas}, {Morrell}, {Phillips}, {Piro}, {Stritzinger}, \& {Suntzeff}}]{Scalzo2019}
{Scalzo}, R.~A., {Parent}, E., {Burns}, C., {et~al.} 2019, \mnras, 483, 628, \dodoi{10.1093/mnras/sty3178}

\bibitem[{{Shara} \& {Hurley}(2002)}]{Shara2002}
{Shara}, M.~M., \& {Hurley}, J.~R. 2002, \apj, 571, 830, \dodoi{10.1086/340062}

\bibitem[{{Shen} {et~al.}(2018){Shen}, {Kasen}, {Miles}, \& {Townsley}}]{Shen2018}
{Shen}, K.~J., {Kasen}, D., {Miles}, B.~J., \& {Townsley}, D.~M. 2018, \apj, 854, 52, \dodoi{10.3847/1538-4357/aaa8de}

\bibitem[{{Smith} {et~al.}(2019){Smith}, {Srivastav}, {McBrien}, {Smartt}, {Gillanders}, {McCormack}, {Denneau}, {Flewelling}, {Heinze}, {Tonry}, {Weiland}, {Rest}, {Clark}, {Fulton}, {O'Neill}, {Young}, \& {Wright}}]{SN2019einDiscovery}
{Smith}, K.~W., {Srivastav}, S., {McBrien}, O., {et~al.} 2019, The Astronomer's Telegram, 12710, 1

\bibitem[{{Soker}(2019)}]{Soker2019}
{Soker}, N. 2019, \nar, 87, 101535, \dodoi{10.1016/j.newar.2020.101535}

\bibitem[{{Taubenberger}(2017)}]{Taubenberger2017}
{Taubenberger}, S. 2017, in Handbook of Supernovae, ed. A.~W. {Alsabti} \& P.~{Murdin}, 317, \dodoi{10.1007/978-3-319-21846-5_37}

\bibitem[{{Timmes} {et~al.}(2003){Timmes}, {Brown}, \& {Truran}}]{Timmes2003}
{Timmes}, F.~X., {Brown}, E.~F., \& {Truran}, J.~W. 2003, \apjl, 590, L83, \dodoi{10.1086/376721}

\bibitem[{{Tully} {et~al.}(2009){Tully}, {Rizzi}, {Shaya}, {Courtois}, {Makarov}, \& {Jacobs}}]{Tully2009}
{Tully}, R.~B., {Rizzi}, L., {Shaya}, E.~J., {et~al.} 2009, \aj, 138, 323, \dodoi{10.1088/0004-6256/138/2/323}

\bibitem[{{Voss} \& {Nelemans}(2012)}]{Voss2012}
{Voss}, R., \& {Nelemans}, G. 2012, \aap, 539, A77, \dodoi{10.1051/0004-6361/201118222}

\bibitem[{{Washabaugh} \& {Bregman}(2013)}]{Wash2013}
{Washabaugh}, P.~C., \& {Bregman}, J.~N. 2013, \apj, 762, 1, \dodoi{10.1088/0004-637X/762/1/1}

\bibitem[{{Xi} {et~al.}(2022){Xi}, {Wang}, {Li}, {Mo}, {Zhang}, {Liu}, {Chen}, {Filippenko}, {Zheng}, {Brink}, {Zhang}, {Sai}, {Ehgamberdiev}, {Mirzaqulov}, \& {Zhang}}]{Gaobo2022}
{Xi}, G., {Wang}, X., {Li}, W., {et~al.} 2022, \mnras, 517, 4098, \dodoi{10.1093/mnras/stac2848}

\end{thebibliography}
\bibliographystyle{aasjournal}



\end{document}